\documentclass{iopart}

\usepackage{epsfig}
\usepackage{float}
\usepackage{graphicx}
\usepackage{color}
\usepackage{cite}

\newcommand{\msd}{\langle x^2 (t)\rangle}
\newcommand{\msdapprox}{ \sigma^2(t)}
\newcommand{\fptd}{f(t)}
\newcommand{\fp}{FPTD}
\newcommand{\kd}{k_D}
\newcommand{\kon}{k_{\rm on}}
\newcommand{\koff}{k_{\rm off}}

\newcommand {\be} {\begin{equation}}
\newcommand {\ee} {\end{equation}}
\newcommand {\bea} {\begin{eqnarray}}
\newcommand {\eea} {\end{eqnarray}}

\begin{document}

\title[First-passage dynamics of obstructed tracer particle diffusion in 1d systems]{First-passage dynamics of obstructed tracer particle diffusion in  one-dimensional systems}

\author{Robin Forsling}
\address{Integrated Science Lab, Department of Physics, Ume{\aa}
University, SE-901 87 Ume{\aa}, Sweden}

\author{Lloyd Sanders}
\address{Department of Astronomy and Theoretical Physics, Lund University,
  S\"olvegatan 14A, SE-223 62 Lund, Sweden}
   
\author{Tobias Ambj\"ornsson}
\address{Department of Astronomy and Theoretical Physics, Lund University,
  S\"olvegatan 14A, SE-223 62 Lund, Sweden}

\author{Ludvig Lizana}
\address{Integrated Science Lab, Department of Physics, Ume{\aa}
University, SE-901 87 Ume{\aa}, Sweden}
\ead{ludvig.lizana@physics.umu.se}

\date{\today}

%

\begin{abstract}
  The standard setup for single-file diffusion is diffusing particles in one
  dimension which cannot overtake each other, where the dynamics of a tracer
  (tagged) particle is of main interest. In this article we generalize this
  system and investigate first-passage properties of a tracer particle when
  flanked by crowder particles which may, besides diffuse, unbind (rebind)
  from (to) the one-dimensional lattice with rates $\koff$ ($\kon$). The
  tracer particle is restricted to diffuse with rate $\kd$ on
  the lattice. Such a model is relevant for the understanding of
  gene regulation where regulatory proteins are searching for specific binding
  sites ona  crowded DNA.  We quantify the first-passage time distribution,
  $\fptd$ ($t$ is time), numerically using the Gillespie algorithm, and
  estimate it analytically. In terms of our key parameter, the unbinding rate
  $\koff$, we study the bridging of two known regimes: (i) when unbinding is
  frequent the particles may effectively pass each other and we recover the
  standard single particle result $f(t)\sim t^{-3/2}$ with a renormalized diffusion constant, (ii)
  when unbinding is rare we recover well-known single-file diffusion
  result $f(t)\sim t^{-7/4}$. The intermediate cases display rich dynamics, with the
  characteristic $\fptd$-peak and the long-time power-law slope both being
  sensitive to $\koff$.
\end{abstract}

\submitto{\NJP}
\maketitle

%
%
\section{Introduction}

In first-passage processes one investigates when a stochastic variable
crosses a given threshold value for the first time
\cite{redner2001guide,condamin2007first}. Such processes are
applicable to many areas, for example two molecules meeting to form a
chemical bond \cite{khairutdinov1996kinetics}, binding of two free
ends of a polymer (so called cyclization) \cite{likthman2006first},
extinction of diseases \cite{khasin2012minimizing}, escape problems
\cite{schuss2007narrow}, arrival times of news and emails
\cite{vazquez2006modeling}, or when a stock market share
crosses a preset market value \cite{bouchaud2000theory}.  Here we
revisit the first-passage problem focusing on a modified version of a
single-file diffusion system where particles not only diffuse and
collide, but also unbind and rebind via a surrounding bulk. Such a
model has relevance for gene regulation where a transcription factor
(protein) searches for its binding site along a crowded DNA
\cite{li2009effects,lomholt2005optimal,kolomeisky2011physics,
  benichou2011intermittent,marcovitz2013obstacles}.

Tracer particle motion in interacting many-body systems is an interesting case
of non-Markovian dynamics \cite{sanders2012first}; memory effects have
profound impact on the system's dynamics. A prototypical example is
single-file diffusion where one studies the tracer particle motion in a system
where particles are confined to one dimension and unable to pass each other
(hardcore repulsion). This restriction leads to correlations between
consecutive jumps of a single particle: if the particle jumps to the right of
its equilibrium position it is on average more likely to move left in the
subsequent jump. Clearly this process is not memoryless and it has been shown
that a tracer particle explores the system under such crowded conditions
subdiffusively (e.g. \cite{harris1965diffusion}); the ensemble averaged tracer
mean squared displacement (MSD) is $\sigma^2(t)\sim \sqrt t$ as a
function of time $t$
\footnote{ Here $\sigma^2(t)=\msd $ where the brackets denote 
    ensemble average. The quantity $x(t)$ is (random) particle position at
    time $t$ with the initial condition $x(t=0)=0$.}.
Even though the microscopic dynamics in a single-file
system is non-Markovian some macroscopic observables exhibit Markovian
behaviour; the centre-of-mass MSD and the dynamic structure factor are two
examples of such slow variables \cite{lizana2009diffusion}. The intuitive
explanation is that it is irrelevant for such macroscopic observables whether
two particles, assumed identical, bounce of each other or pass through each
other.
Single-file diffusion, has attracted great interest for at last 50
years and the body of work is substantial, both on the theoretical
\cite{harris1965diffusion, levitt1973dynamics, kehr1981diffusion,
  percus1974anomalous, hahn1999propagator, van1985mean,
  schutz1997exact, burlatsky1996motion, alexander1978diffusion,
  jara2006nonequilibrium, arratia1983motion, kollmann2003single,
  lizana2008single, taloni2008langevin, barkai2009theory,
  lomholt2011dissimilar,rodenbeck1998calculating}, and experimental
\cite{wei2000single,lin2005random, kukla1996nmr,
  meersmann2000exploring, hodgkin1955potassium} side.

In this paper we are interested in the first-passage time density (\fp), $\fptd$,
for a tracer particle in a single-file system where all particles, except for
the tracer, are allowed to detach from the line and rebind at a random
location. The detachment rate, $\koff$, is the main parameter of consideration
since it allows us to study the transition between two known regimes:
\begin{itemize}
\item[(1)] {\it Single-particle regime:} For large unbinding rates particles may pass each other
  and the no-passing condition is violated. The tracer particle is
  therefore diffusing as if it was free \cite{redner2001guide}.
\item[(2)] {\it Single-file regime:} When the unbinding rate is small (compared to the
  diffusion rate) the non-passing condition is only weakly
  broken. Here we expect to recover the known \fp~for single-file
  diffusion \cite{sanders2012first,molchan1999maximum}.
\end{itemize}
We also propose a closed form expression for $\fptd$ based on a Markovian
assumption which allow us to quantify the increase in
  non-Markovian effects as $\koff$ is lowered.

The organization of the paper is as follows: In Section \ref{sec:model} we
define our model. In Section \ref{sec:analyt} we provide a simple analytical
treatment to increase the understanding of our numerical results which are
presented and discussed in Section \ref{sec:results}. Details
of the numerics are placed in  \ref{sec:numerics}. Finally, in Section
\ref{sec:conclusion} we conclude our work and discuss future directions.

%
%
\section{The model}\label{sec:model}

\begin{figure}
\center
\includegraphics[width=10cm]{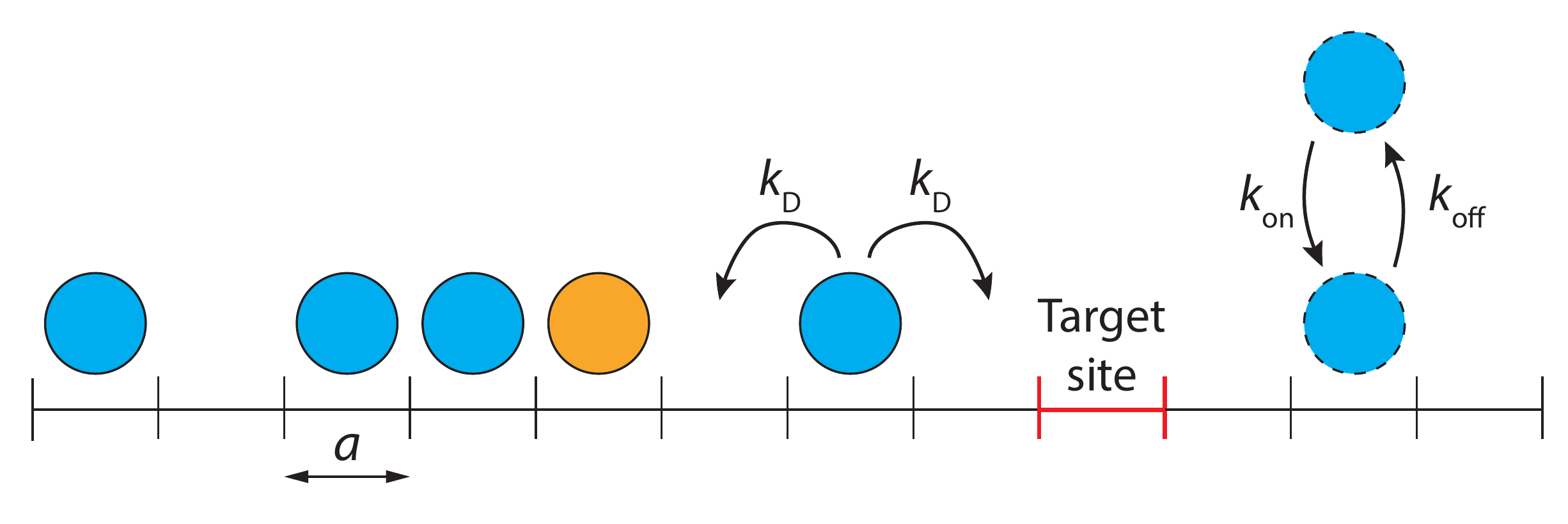}
\caption{Schematic illustration of our model. Particles are diffusing with
  rate $\kd$ on a one dimensional lattice with lattice spacing $a$.  All
  particles, except for the tagged one (orange), may also unbind and rebind
 at a random lattice site
 with rates $\koff$ and $\kon$, respectively. The tagged particle cannot
  leave the line ($\kon=\koff=0$) when searching for the target site. }
\label{fig:model}
\end{figure}

Consider a one dimensional lattice on which crowder particles and the tracer
particle diffuse (Figure \ref{fig:model}). The crowder particles can in addition to  diffuse, unbind and rebind to the lattice. 
Rebinding occurs to a random unoccupied lattice site. The lattice
constant is denoted $a$ and the diffusion rate $\kd$ is assumed equal in both
direction and for all particles
%
\footnote{There is nothing preventing us from letting
  $\kd$ be different for all particles, similar to \cite{lomholt2011dissimilar}.}.  
Double occupancy is forbidden and a particle cannot overtake a flanking
neighbor (single-file condition). Binding and unbinding dynamics of crowders
are characterized by the rates $\kon$ and $\koff$; they are chosen such
that the particle concentration on the line is in equilibrium with the
bulk. We keep the the average filling fraction at 25\% during the
simulations. The ends of the lattice are reflecting but the number of lattice
sites chosen sufficiently large such that boundary effects do not affect our
results
\footnote{Particles in the bulk rebind to a random unoccupied lattice
site. This means that we model the bulk in an effective way and neglect that
rebinding close to the unbinding site is more probable \cite{kolomeisky2011physics}. This would be particularly
important if the surrounding bulk was crowded as is the likely scenario in the
interior of a living cell.}.

To consider the first-passage properties of the tagged particle in the 
aforementioned system, we introduce a perfectly absorbing target site (see Figure \ref{fig:model}).
This site reacts only with the tracer and remains inert for all other particles.
In this scenario we seek to understand the FPTD of the tracer to the target
site as a function of $\koff$.

We implemented the model described above using the Gillespie algorithm;
details are deferred to \ref{sec:numerics}. Briefly, in a simulation we
initially place the particles randomly (thermal initial condition) where the
initial position of the tracer particle is 70 sites away from the target site. Also, due to recent
interest in non-thermal initial conditions in single-file diffusion
\cite{lizana2014single,leibovich2013everlasting} we in addition investigate
the case when the particles are placed equidistantly for $\koff=0$.
We run the simulation until the tagged particle hits the absorbing target and
make a record of the absorption time. From many such runs we then determine
the ensemble averaged FPTD (normalized histogram of absorption times) as a
function of our key parameter $\koff$.  In this way we interpolate between the
single-file regime ($\koff \ll \kd$) to the unobstructed single particle
regime ($\koff \gg \kd$).

%
%
\section{Analytical estimates}\label{sec:analyt}

In this section we provide analytical estimates to corroborate the numerical
results in the next section. First, we put bounds on the tracer's MSD, 
denoted by $\msdapprox$, in the absence of the absorbing
site. Second, we also put bounds on the \fp~by discussing established results. Third, we relate $\msdapprox$ to the first-passage time density, based on a Markovian approximation. These estimates allow us to quantify
non-Markovian effects in the \fp~simulations in the Results section.

\subsection{Bounds and estimates for $\msdapprox$}

Let us now provide estimates for the MSD for the process
as considered here and depicted in Figure \ref{fig:model}. Clearly the obstructed diffusion process  can never be faster than if there were no surrounding
particles at all. Therefore an upper bound on $\msdapprox$ is
the unobstructed (single particle) diffusion result
\begin{equation}\label{eq:upper}
\hat \sigma^2_{\rm SP}(t)=2D_{\rm SP}t^{2H_{\rm SP}}
\end{equation} 
with a diffusion constant $D_{\rm SP}=2\kd a^2$. The, so called,
  Hurst exponent $H=H_{\rm SP}=1/2$ relates the distance ``explored'' by the
  particle to time, $t$.

The lower bound on $\msdapprox$ is reached when no
particles unbind at all. This is the single-file (SFD) regime where
\begin{equation}\label{eq:lower}
\hat \sigma^2_{\rm SFD}(t)= \gamma \frac{(1-\rho a )}{\rho}
\sqrt{\frac{4D_{\rm SP}}{\pi}} t^{2H_{\rm SFD}}
\end{equation}
with a Hurst exponent $H=H_{\rm SFD}=1/4$ and $\rho$ is the particle
density. We point out that tracer particle motion in a
  single-file system is dominated by long-time memory (non-Markovian) effects,
   manifested through a subdiffuse Hurst exponent ($H<1/2$).  For
equidistant initial conditions $\gamma=1/\sqrt{2}$
\cite{lomholt2011dissimilar,leibovich2013everlasting,lizana2010foundation},
whereas $\gamma=1$ for thermal initial conditions. In our single-file
simulations we address both cases.



\subsection{Established results  for the  \fp}

Here we give an explicit expression for the \fp~ which is valid
for the single-particle and single-file regimes.  For a freely
diffusing particle the first-passage time density is
well-known \cite{redner2001guide}. Also, rather recently, extensive
simulations led to a proposed form \cite{sanders2012first} for
single-file diffusion. The results for these two cases are contained in
the following expression \cite{sanders2012first}
\be\label{eq:f}
f(t) = C \frac{1}{t} \left[\frac{\Delta x}{\sqrt{2}\ \sigma(t)} \right]^{(1-H)/H}\exp\left[-\frac{\Delta x^2}{2\sigma^2(t)} \right],
\ee
with normalization constant $C=2H/\Gamma[(1-H)/(2H)]$,
where $\Gamma(z)$ is the gamma-function, and $\Delta x$ is distance to target at $t=0$.
The single particle result is found when $H=1/2$, and the single file
result is $H=1/4$. The mean square displacements, $\sigma^2(t)$, are
given in equations (\ref{eq:upper}) and (\ref{eq:lower}) respectively.  The
functional form above provides a long time tail $t^{-3/2}$ for a single
particle \cite{redner2001guide} as it should, and is in agreement with
the exact asymptotic result $t^{-7/4}$ \cite{molchan1999maximum} for single file
diffusion. 

\subsection{Markovian approximation for the \fp}

It is unclear to what extent equation (\ref{eq:f}) holds when
$0<\koff<\infty$, i.e., when we are in neither the single-particle regime
($\koff=\infty$) nor in the single-file regime ($\koff=0$). In
  particular, direct application of equation (\ref{eq:f}) is complicated by
  the fact that for general $\koff$, the Hurst exponent $H$, appearing in
  equation (\ref{eq:f}), does not take a universally valid value for all times
  (see MSD simulations in the next section). Below we therefore provide a
useful form for the \fp~ which will allow us to quantify non-Markovian effects
in our \fp~ simulations (see Section \ref{sec:results}). To that end, we
define:
\begin{equation}\label{eq:1}
 Q(t) = \frac{|\Delta x|}{\sqrt{2 \msdapprox }},   
\end{equation}
where $\Delta x$, as before, is the initial distance to the target. In
terms of $Q(t)$ we express the \fp~ for standard diffusion (Markovian
process) as \cite{redner2001guide}
\begin{equation}\label{eq:2}
 \fptd = A \frac{Q(t)}{t} \exp[-Q^2(t)],      
\end{equation}
with normalization constant $ A= 1/\int_0^\infty [Q(t)/t] \times
\exp[-Q^2(t)]dt $. If $\msdapprox \propto t$, then $A=1/\sqrt \pi$.  Equation
(\ref{eq:2}) can be obtained by putting $H=1/2$ in equation (\ref{eq:f}) or
from using the method of images \cite{redner2001guide}.  Method of images is
an appealingly simple way to solve first-passage problems to
  perfectly absorbing boundaries, but, it has limited applicability for
  non-Markovian dynamics \cite{sanders2012first, jeon2011first}. Nevertheless,
  this method been used previously for a similar setup to ours
  \cite{li2009effects}.

When $\koff \rightarrow \infty$ our system is Markovian whereas in the
opposite limit $\koff \rightarrow 0$ it is highly non-Markovian.  Since
equation (\ref{eq:2}) is based on a Markovian assumption it cannot be used to
calculate the correct $\fptd$ for single-file diffusion, only to quantify
deviations from non-Markovian dynamics. To see this explicitly, using
$\msdapprox\propto \sqrt t$ in equation (\ref{eq:2}), yields $\fptd \sim
t^{-5/4}$, rather than the exact result $\fptd\sim t^{-7/4}$
\cite{sanders2012first,molchan1999maximum}.  If we instead use
$\msdapprox\propto t$ we obtain the proper $\fptd\sim t^{-3/2}$
for standard diffusion to an absorbing boundary.

In the next section we utilize equations (\ref{eq:1}) and (\ref{eq:2}) in the
following way: first, we run simulations to numerically determine the MSD for
different parameter sets. Subsequently we use the raw data from those
simulations as $\msdapprox $ which we substitute into equations (\ref{eq:1})
and (\ref{eq:2}). In this way we calculate an approximation for the
\fp. Finally, we perform explicit \fp~simulations and compare those to the
Markovian approximation for the FPTD given by equations
  (\ref{eq:1}) and (\ref{eq:2}).

%
%

\section{Results}\label{sec:results}

In this Section we present the results of stochastic simulations
(Gillespie algorithm) of the model outlined in section
\ref{sec:model}.  The main quantity of interest is the \fp~ to a 
  perfectly absorbing
target site for different unbinding rates $\koff$. But first, we
characterize our system in terms of the tracer particle MSD. The MSD
simulations are subsequently used to test for non-Markovian deviations
in our first-passage simulations using equation (\ref{eq:2}).

\begin{figure}
\center
\includegraphics[width=10cm]{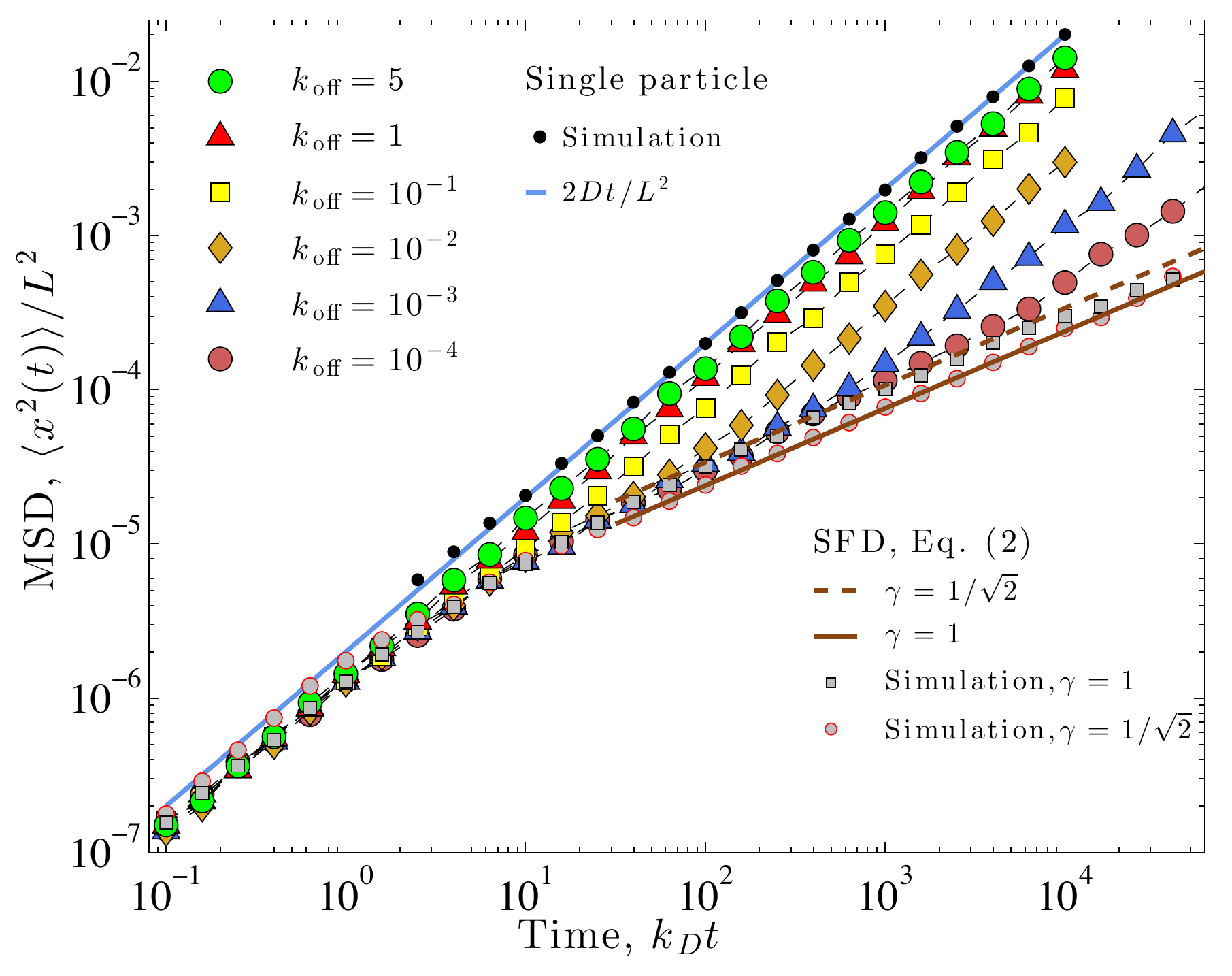}
\caption{Tracer particle MSD vs. time in an
  obstructed diffusion system, for different off-rates. As the off-rate is
  increased the MSD of the tracer particle approaches the single-particle result $2Dt$ (solid light blue line). In all simulations we used: diffusion rate $\kd= 1$, lattice spacing $a=1$ and number of lattice sites $L=1001$. The data was ensemble averaged over $2000$ simulation runs. The unbinding rates are listed in the figure
  legend, and the binding rates were chosen to maintain the average filling fraction 0.25 
  (see \ref{sec:numerics}). 
}
\label{fig:msd}
\end{figure}

In Figure \ref{fig:msd} we show the tracer particle MSD as a function
of time for different $\koff$ (keeping all other parameters fixed). We
notice that as $\koff$ is increased the MSD is increased; the larger
the unbinding rate, the less obstructed the tracer particle motion
is. All MSD simulations are within the upper and lower bounds as
provided by the single-particle limit [equation (\ref{eq:upper})] and
the single-file diffusion limit [equation (\ref{eq:lower})] as
expected. For small $\koff$ the system behaves just as a single-file
system up to a cross-over time $\tau^*$, after which the MSD is linear
in time.  For the case that $\koff$ is the rate limiting parameter, we
simply estimate the crossover time as the (average) time for an
unbinding event to take place, i.e., $\tau^*\approx1/\koff$; this prediction for $\tau^*$ is in agreement with our simulation results for $\koff \le 0.1$. 

  Figure \ref{fig:msd} also shows the MSD for single-file diffusion when
  particles placed equidistantly at $t=0$ (non-thermal initial condition). The
  result shows that the asymptotic form of the MSD is well described by
  equation (\ref{eq:lower}), with $\gamma =1/\sqrt 2$. We are the first to
  numerically verify this result which, up to now, only was known on analytical
  grounds
  \cite{lomholt2011dissimilar,leibovich2013everlasting,lizana2010foundation}.

\begin{figure}
\center
\includegraphics[width=10cm]{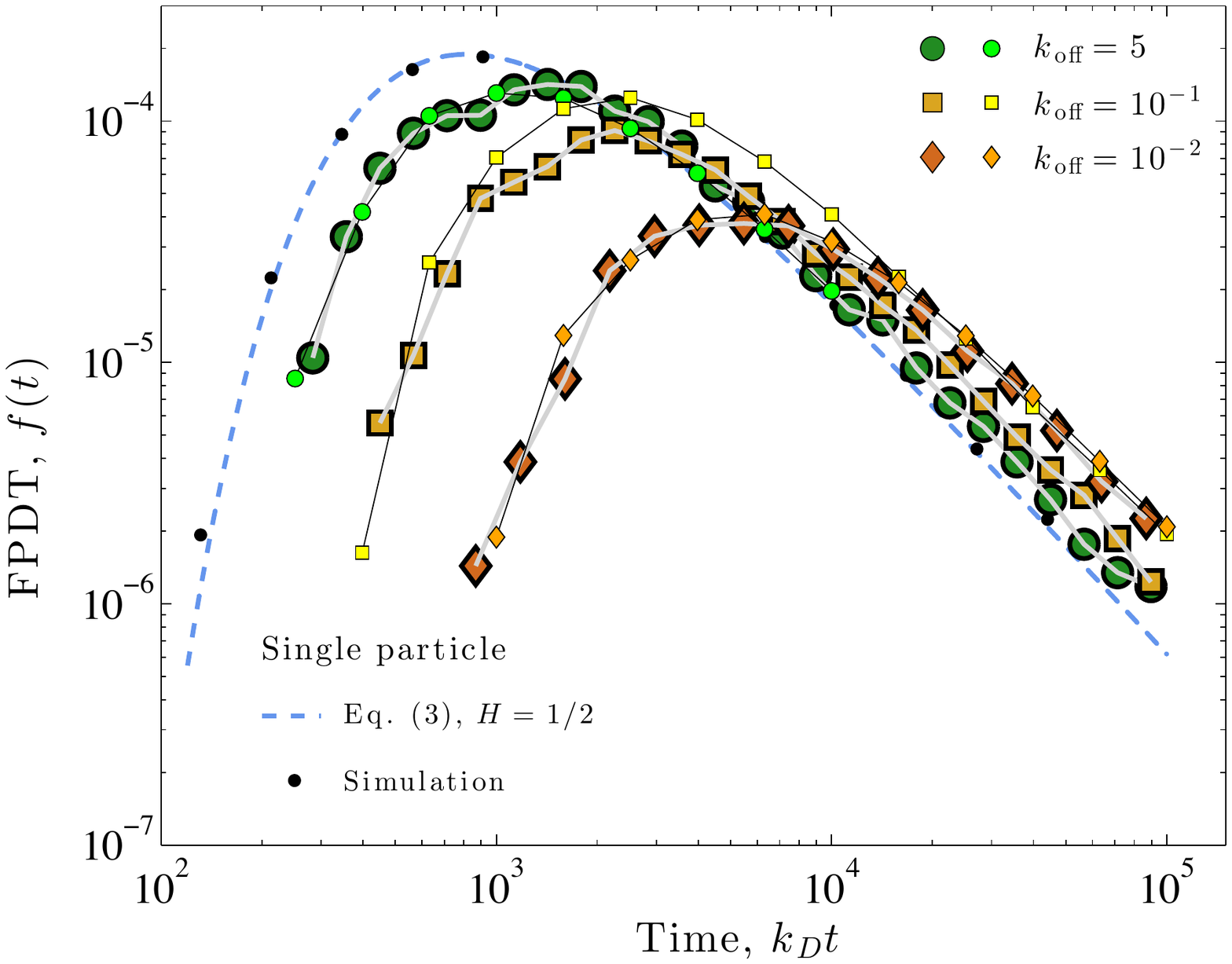}
\includegraphics[width=10cm]{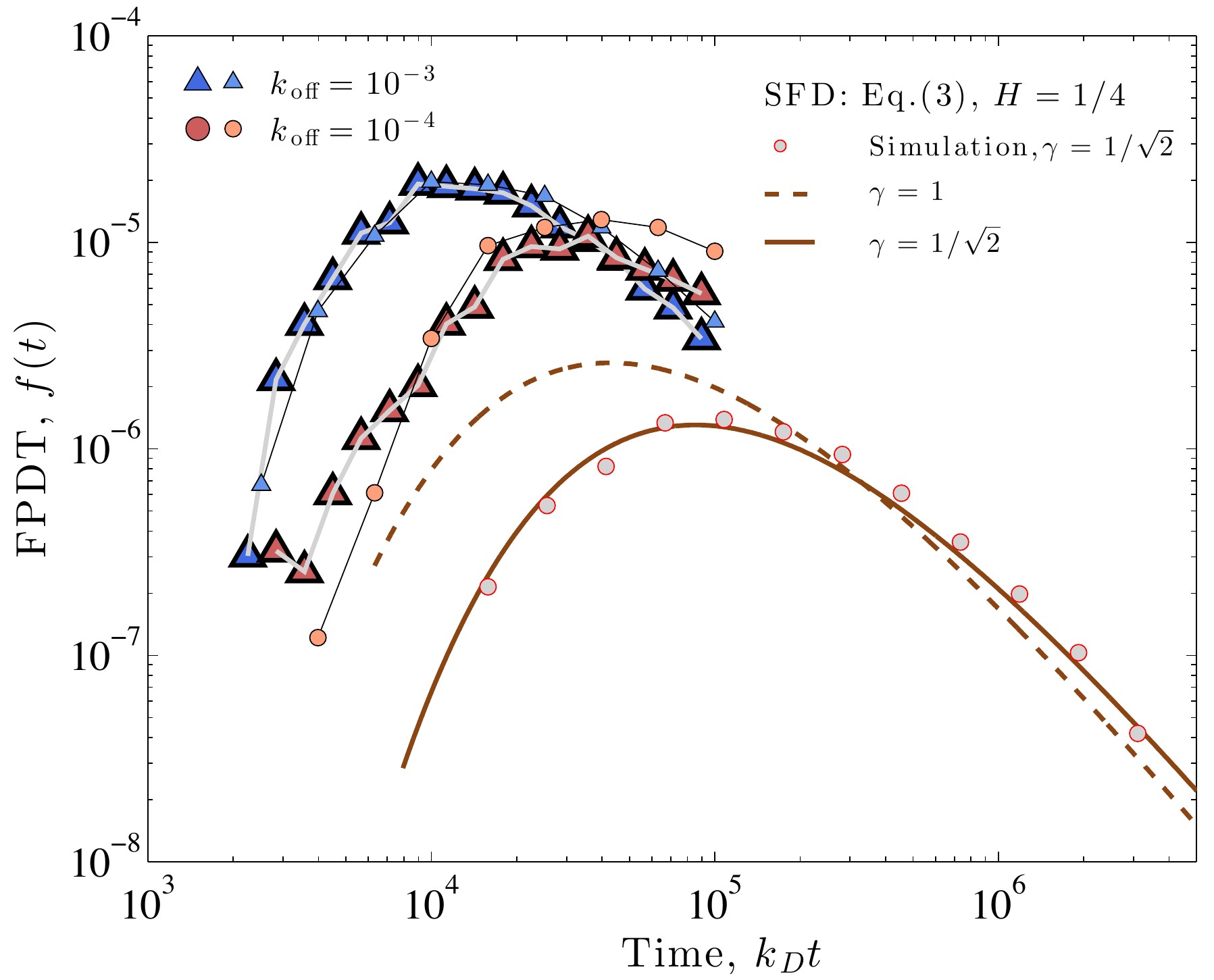}
\caption{First-passage time density of a tracer particle in an obstructed
  diffusion system for (top) large unbinding rates $\koff$ and
    (bottom) small $\koff$. Large symbols denote normalized
  first-passage time histograms when the target (placed in the middle of the
  lattice) was initially 70 lattice sites away from the tracer particle's
  initial position.  The same parameters as listed in Figure \ref{fig:msd}
  were used also here ($\kd=1$, $a=1$ and $L=1001$).  The first-passage data
  is ensemble averaged over $9.6\cdot10^3$ ($\koff=10^{-4}$ and
  $\koff=10^{-3}$), $10^4$ ($\koff=0.01$ and $\koff=0.1$), $4.8\cdot 10^3$
  ($\koff=5)$, $10^3$ ($\koff=0$, SFD), and $5\cdot 10^5$ ($\koff=\infty$,
  single particle) simulation runs. Small symbols show the performance of
  the Markovian approximation, equation (\ref{eq:2}), where we
  used the MSD simulations as $\sigma(t)$ in
  equation(\ref{eq:1}). Notice that as $\koff$ is lowered
    non-Markovian effects become more pronounced.}
  \label{fig:fptd}
\end{figure}

Figure \ref{fig:fptd} shows simulation results for the probability density of
first-passage times $\fptd$ for different $\koff$ (small symbols).  For
clarity, the top panel shows the results for large $\koff$ whereas the bottom
panel holds the results for small $\koff$.  We notice that as $\koff$
increases $\fptd$ approaches that for a single particle (top panel). As
$\koff$ gets smaller the maximum of the curve is shifted towards longer times
and the power-law exponent of tail decreases. This agrees with
the fact that longtime exponent should change from -3/2 to -7/4
as $\koff$ decreases.

We also compare simulations to analytical approximations in
  figure \ref{fig:fptd}. First, we find that for the case of a single particle
  (top panel), equation (\ref{eq:f}) with $H=1/2$ agrees well with simulations
  as it should. For the case of single-file diffusion with equidistant initial
  conditons (bottom panel), $\gamma=1/\sqrt{2}$, we find good agreement
  between simulations and equation (\ref{eq:f}) with $H=1/4$ (the case
  $\gamma=1$ was considered extensively in \cite{sanders2012first}). Our
study therefore extends the applicability of this expression to a more general
setting than considered previously \cite{sanders2012first}, where only thermal
initial conditions were employed. Equation (\ref{eq:f}) can, however, not be
used for general off rates; as noted earlier, for general $\koff$
  it is not straightforward to choose a proper Hurst exponent that
  characterises our system for all times. Instead we compare our simulations
to the Markovian approximation (\ref{eq:2}) (large symbols), with $\msdapprox$
extracted from Figure \ref{fig:msd}. We find that for all large $\koff$ values
considered (top panel), the discrepancy between the simulations
and equation (\ref{eq:2}) are small. This indicates that non-Markovian effects
are small in this regime.  On the other hand, when $\koff$ gets smaller
(bottom panel) the deviations from equation (\ref{eq:2})
grow and the non-Markovian nature of the problem becomes
increasingly prominent.

%
%
\section{Concluding remarks and outlook}\label{sec:conclusion}

We studied first-passage statistics of a tracer particle in a single-file
diffusion system which can exchange particles with a surrounding bulk. We
focused mainly on the probability density of first-passage times, $\fptd$,
which we quantified numerically and estimated analytically. We distinguished
two limiting behaviors as a function of our main parameter $\koff$. When the
unbinding rate is small ($\koff \ll \kd$) we recover single-file results. When
the unbinding is large ($\koff \gg \kd$) the no-passing condition is
effectively violated and we recover the result for a single particle. The
transient function connecting these two regimes is non-trivial and was
characterized in terms of $\koff$. For the single-particle and the single-file
cases our simulations further established the validity of equation
(\ref{eq:f}). For general off-rates, equation~(\ref{eq:2}) was used as a
quantifier of non-Markovian effects, and we find that such effects
  become more prominent as $\koff$ is lowered.

Single-file diffusion is a useful model system to study effects of
crowding in diverse (bio)physical media. It has received much of
attention over the last five decades partly due to its analytical
tractability. However, in its original formulation particles are under
no conditions allowed to pass each other, which in real systems rarely
is the state of affairs. One example is target-finding problems on DNA
where most studies known to the authors omit crowding, see for example
\cite{li2009effects,lomholt2005optimal,benichou2011intermittent}. 
It is therefore a challenge to extend current
knowledge on single-file diffusion to reach a wider applicability.  We
hope that our work will inspire future progress in this direction.

%
%

\ack

LL acknowledges the Knut and Alice Wallenberg foundation  and the Swedish
Research Council (VR), grant no. 2012-4526, for
financial support. TA is grateful to VR for funding (grant no. 2009-2924).

\appendix
\section{Numerical implementation}\label{sec:numerics}

The model, as depicted in Figure \ref{fig:model}, is implemented
numerically using Gillespie's algorithm
\cite{gillespie1976general}. Each particle on the lattice is given
rates corresponding to its allowed actions: a rate of jumping left or
right ($\kd$) and a rate for unbinding ($\koff$). The binding rate
$\kon$ is adjusted during the simulation such that the average number
of occupied lattice sites $\rho$ is kept fixed. Detailed balance gives
\be \label{eq:kon}
\kon = \koff \bar{\rho}(t)/\varrho_{\rm bulk},  
\ee
where $\bar{\rho}(t)$ is the instantaneous particle concentration of
the lattice (number of particles divided by the system's length), and
$\varrho_{\rm bulk}$ is the concentration in the bulk. Relation
(\ref{eq:kon}) simply means that the bulk acts as a particle
reservoir. Low copy number fluctuations in the surrounding volume can,
in this way, not be captured with our implementation.  The no-passing
condition between particles is implemented by setting the jump rate to
zero in the direction of a flanking particle. The tracer particle is
different from all others since the binding and unbinding rates are
put to zero ($\koff=\kon=0$) throughout the simulation. Its diffusive
motion is otherwise the same.  In the simulations we used $\rho=0.25$,
$\kd = 1$, $a=1$, $\Delta x = 70$, and used $\koff$ as the free parameter
in the simulation.

Below follows a few technical details on the implementation. First, we
define a binary occupation vector $\mathbf X$ with the same length as
the number of lattice sites $L$. A '0' ('1') entry indicates vacant
(occupied) sites. Second, we define a rate vector $\mathbf K$
containing all particle rates.  This vector contains $3L+1$ entries
ordered in the following way. If for example $X_j=1$ and $X_{j\pm1}=0$
then we put $K_{2j-1}=\kd$, $K_{2j}=\kd$ for jumping, and
$K_{j+2L}=\koff$ for unbinding. If, on the other hand, a neighbouring
site was occupied, say $X_{j+1}=1$, then we must put $K_{2j}=0$ (and
 $K_{2j+1}=0$ for the particle at site $j+1$), such that the particles
are not allowed to pass each other.  The rate for adding a new
particle to the lattice (i.e. binding) is for convenience placed in
the last element $K_{3L+1}$. Below is a schematic outline of our
algorithm:
\begin{enumerate}
\item Place the tracer 70 sites away from the target. The
    remaining particles are placed randomly to the left and right of the
    tracer particle. Reset the simulation time~$t_{\rm tot}=0$. For the
    single-file case with $\gamma=1/\sqrt{2}$ the particle are placed
    equidistantly at the start of the simulation. 
\item Assign to each particle their corresponding rates, sum all
  elements in $\mathbf K$, $k_{\rm tot}=\sum_i K_i$.
\item Draw two uniformly distributed random numbers $r_1$ and $r_2$
  between 0 and 1.
\item From $r_1$, calculate time until next event $\tau$, according to
  $\tau=-(1/k_{\rm tot})\ln r_1$, and update time $t_{\rm
    tot}\rightarrow t_{\rm tot}+\tau$.
\item From $r_2$, determine the particle event by finding the index
  $\mu$ in $\mathbf K$ that satisfies
  \[ \sum_{i=0}^{\mu-1}K_{i} <k_{\rm tot} r_2 \leq \sum_{i=0}^{\mu}
  K_i. \] Each $\mu$ corresponds to one particle doing one of its
  allowed actions both of which are inferred from the index $\mu$ of
  $\mathbf K$. If a particle is at site $j$ ($X_j=1$) and unbinds then
  we simply put $X_j=0$.  Or, if the same particle jumps right, then
  we set $X_j=0$ and $X_{j+1}=1$.
\item Return to step 2 and repeat until the tracer particle has
  reached its target. The ensemble average is obtained by repeating
  the steps 1-6 many $(\sim 10^3-10^4)$ times.
\end{enumerate}
The absorbing target is omitted in step 6 when we are only
interested in the tracer particle's MSD.

What we describe above is a generalization of the so-called direct
method for single-file diffusion systems. For the single-file
simulations ($\koff=0$) displayed herein we employ a
  computationally improved approach, the trial and error method, which
  is based on the idea that the sum of $\mathbf K$ is carried out only
  once \cite{ambjornsson2008single}. This method cannot be
 directly applied for $\koff  \neq 0$ since the sum over
  $\mathbf K$ must be recalculated every time a new particle
  associates or dissociates.

\section*{References}


\end{document}